\documentclass[aps,prd,onecolumn,groupedaddress,nofootinbib,showpacs,showkeys,amssymb,11pt]{revtex4}
\usepackage[a4paper,left=25mm,right=25mm,top=25mm,bottom=25mm]{geometry}
\usepackage{epsfig} 
\usepackage{amsmath} 
\usepackage{amsfonts}
\usepackage{amssymb} 
\usepackage{graphicx} 
\usepackage{colordvi}
\def\be{\begin{equation}}
\def\ee{\end{equation}}
\def\bea{\begin{eqnarray}}
\def\eea{\end{eqnarray}}

\def\d{\delta}

\def\pa{\partial}

\let\alp=\alpha

\renewcommand{\epsilon}{\varepsilon}

\def\beqa{\begin{eqnarray}}
\def\eeqa{\end{eqnarray}}
\def\beq{\begin{equation}}
\def\eeq{\end{equation}}

\def\pa{\partial}

\let\alp=\alpha

\renewcommand{\epsilon}{\varepsilon}

\def\pa{\partial}

\def\alp{\alpha}

\def\bq{{\bf q}}
\def\bqd{{\bf \dot{q}}}
\def\qd{\dot{q}}

\def\d{\partial}

\def\eqdef{\buildrel {\rm def} \over =}
\def\({\left(}    \def\){\right)}

\def\data{\the\day-\the\month-\the\year}

\def\frac#1#2{{#1 \over #2}}

\def\phi{\varphi}

\def\~{\approx}

\usepackage{latexsym,float}
\usepackage{epsfig,graphics}
\usepackage{epstopdf}
\usepackage{amssymb}
\usepackage{amsmath}
\usepackage{verbatim}
\usepackage{multirow}
\usepackage{color}
\usepackage{tikz}
\usepackage[utf8]{inputenc}
\usepackage{commath}
\usepackage{braket}
\usepackage{amssymb}
\usepackage{lineno}
\usepackage{graphicx}

\def\gtwid{\mathrel{\raise.3ex\hbox{$>$\kern-.75em\lower1ex\hbox{$\sim$}}}}
\def\ltwid{\mathrel{\raise.3ex\hbox{$<$\kern-.75em\lower1ex\hbox{$\sim$}}}}
\def\square{\kern1pt\vbox{\hrule height 1.2pt\hbox{\vrule width 1.2pt\hskip 3pt
			\vbox{\vskip 6pt}\hskip 3pt\vrule width 0.6pt}\hrule height 0.6pt}\kern1pt}
\begin{document}

\title{Noether symmetry in the Nash theory of gravity}
        
\author{Phongpichit Channuie}
\email{channuie@gmail.com}
\affiliation{ School of Science, Walailak University, Thasala, \\Nakhon Si Thammarat, 80160, Thailand}
\author{Davood Momeni} \email{davood@squ.edu.om}
 \affiliation{{Department of Physics, College of Science, Sultan Qaboos University,
		\\P.O. Box 36, P.C. 123,  Muscat, Sultanate of Oman}}
        
\author{Mudhahir Al Ajmi} \email{mudhahir@squ.edu.om}
\affiliation{{Department of Physics, College of Science, Sultan Qaboos University,
		\\P.O. Box 36, P.C. 123,  Muscat, Sultanate of Oman}}

\begin{abstract}

This paper deals with the study of Bianchi type-I universe in the context of Nash gravity by using the Noether symmetry approach. We shortly revisit the Nash theory of gravity. We make a short recap of the Noether symmetry approach and consider the geometry for Bianchi-type I model. We obtain the exact general solutions of the theory inherently exhibited by the Noether symmetry. We also examine the cosmological implications of the model by discussing the two cases of viable scenarios. Surprisingly, we find that the predictions are nicely compatible with those of the $\Lambda$CDM model.
 
\end{abstract}

\date{\today}

\maketitle

\section{Introduction}
Several cosmological observations show that the observable
universe is undergoing a phase of accelerated expansion \cite{perl}.  Regarding the late-time cosmic acceleration,
there are at least two promising explanations, to date. The first one is to introduce the dark energy component in the universe \cite{sahni}. However, the dark energy sector of the universe remains still unknown. Conversely, the second popular approach is to interpret this phenomenon by using a purely geometrical picture. The later is well known as the modified gravity. Modified theories of gravity have received more attention lately due to numerous motivations ranging from high-energy physics, cosmology and astrophysics \cite{od}.

The modified theories of gravity can be in principle achieved from different contexts. One of the earlier modifications to Einstein's general relativity was known as the Brans-Dicke gravity. In addition to a gravitational
sector, this theory introduced a dynamical scalar field to represent a variable gravitational constant \cite{dicke}. Later it was found that the authors of Ref.\cite{sb} studied a scalar-tensor theory of gravity in which the metric is coupled to a scalar field. Regarding the work of Ref.\cite{sb}, a `missing-mass problem' can be successfully described. Moreover, this approach can be applied to the Bianchi cosmological models.

Another simplest modification to the standard general relativity is the $f(R)$ theories of gravity in which the Lagrangian density $f$ is an arbitrary function of the scalar curvature $R$ \cite{  Bergmann:1968ve,Buchdahl:1983zz}. Among numerous alternatives, these theories include higher order curvature invariants, see reviews on $f(R)$ theories \cite{Sotiriou:2008rp,DeFelice:2010aj}. In recent years, a new stimulus for this study leads to a number of interesting results. Notice that the model with $f(R)=R+\alpha R^{2}\,(\alpha > 0)$ can lead to the accelerated expansion of the Universe because of the presence of the $\alpha R^{2}$ term. This particular case is the first model that can describe cosmic inflation proposed by Starobinsky \cite{Starobinsky:1980te}. There exists another different class of the modified gravity theory, called MOG, which can alternatively explain the flat rotation curve of galaxies without invoking cold dark matter particles \cite{Moffat:2005si} (see also recent examination \cite{Roshan:2014mqa}). Likewise, John Nash has developed an alternative theory to the Einstein’s theory. The theory has been proven to be formally divergence free and considered to be of interest in constructing  theories of quantum gravity.

In order to quantify the exact solutions, it has been proven that the Noether symmetry technique proved to be very useful not only to fix physically viable cosmological models with respect to the conserved quantities, but also to reduce dynamics and achieve exact solutions. Moreover, the existence of Noether symetries plays crucial roles when studying quantum cosmology \cite{SC}. In addition, the Noether symmetry approach has been employed to various cosmological scenarios so far  including the $f(T)$ gravity \cite{Channuie:2017txg}, the $f(R)$ gravity \cite{j}, the alpha-attractors \cite{Kaewkhao:2017evn}, and others cosmological scenarios, e.g. \cite{cap3,Momeni:2015gka,Momeni:2014iua,Aslam:2013pga,Aslam:2012tj,Jamil:2012fs,Jamil:2012zm,Jamil:2011pv,Bahamonde:2017sdo}. Moreover, the Noether symmetry approach has been also utilized to study the Bianchi models \cite{Capozziello,camci} in order to obtain the exact solutions for potential functions, scalar field and the scale factors.

In this paper, we examine a Bianchi type I spacetime in the
framework of Nash Gravity by using the Noether symmetry approach. The structure of the paper is as follows: In Sec.(\ref{II}), we make a short recap of the Nash theory of gravity. Here we display the gravitational equations of the Nash theory in vacuum. In Sec.(\ref{III}), we revisit the Noether symmetry approach and consider the geometry for Bianchi-type I model. In Sec.(\ref{IV}), we study exact solutions exhibited from the Noether symmetry. We also examine in Sec.(\ref{v}) the cosmological implications of the model by discussing the two cases of viable scenarios. Finally, we conclude our findings in the last section.

\section{A short recap of Nash's theory for gravity}
\label{II}
During his lifetime, J. Nash tried once to develop an alternative theory of gravity. The theory is obtained as a modification of GR and can be considered as an alteranative to make GR renormalizable \cite{Nash}. More recent works are presented in \cite{Lake:2017uic,Aadne:2017oba}. Let us have a look on the Nash gravity at action level:
\begin{equation}
{\cal S}=\int d^{4}x {\cal L}=\int d^{4}x\sqrt{-g}\Big(2R^{\mu\nu}R_{\mu\nu}-R^2\Big)\,. \label{action}
\end{equation}
It is remarked that the general class of Lagrangians including the one written above has been considered to be of interest in attempting to develop theories of quantum gravity. Using the above action, gravitational field equations are directly derived by taking into account the metric $g^{\mu\nu}$ as a dynamical field to yield
\begin{equation}
\Box G^{\mu\nu} + G^{\alpha\beta}\Big(2R_{\alpha}^{\,\,\mu}R_{\beta}^{\,\,\nu} - \frac{1}{2}g^{\mu\nu}R_{\alpha\beta}\Big)=0,\label{eq}
\end{equation}
where $\Box$ is the d’Alembertian operator, and $G^{\mu\nu}=R^{\mu\nu}-\frac{1}{2}g^{\mu\nu}R$. It was demonstrated that all Ricci flat solutions (de Sitter like metrics) in Nash theory coincide to GR solutions in four dimensions. Stelle showed that such actions were re-normalizable \cite{Stelle:1976gc}. In the flat FLRW background, it is still illustrative to study cosmological solutions; see also a previous work \cite{Lake:2017uic}. Due to the issue of stability, it will be motivating to examine cosmological solutions for homogeneous but anistropic universe.

Recently, Strumia and Salvio extended Stelle's work to their "agravity" theory \cite{Salvio:2014soa}. They were motivated to consider this action because of a principle of "classical scale invariance." This implies a renormalizable quantum gravity theory where the graviton kinetic term has 4 derivatives, and can be reinterpreted as graviton minus an anti-graviton. In addition, within agravity, they found that inflation is a generic phenomenon.

\section{Noether symmetry \& Bianchi-type I Universe}
\label{III}

From the viewpoint of the Noether symmetry approach, exact solutions of dynamical system, that is a point-like Lagrangian, can be achieved by selecting cyclic variables. In principle, the corresponding dynamical system
can be completely integrated and the potential exhibited by the symmetry can be exactly obtained.  Based on the Noether theorem, if there exists a vector field ${\bf X}$, for which the Lie derivative of a given Lagrangian ${\cal L}$ vanishes, i.e. $L_{{\bf X}}{\cal L} = {\bf X}{\cal L} = 0$, the Lagrangian admits a Noether symmetry and thus yields a conserved current. Note that the Noether symmetry dictates a local transformation. A point transformation defined by $Q^{i}=Q^{i}({\bq})$ can in principle depend on one (or multiple) parameter(s). In this situation, the vector field ${\bf X}$ takes the form
\beq \label{05} 
{\bf X}=\alp^{i}({\bq})\frac{\pa}{\pa q^{i}}+
\left(\frac{d}{d\lambda}\alp^{i}({\bq})\right)\frac{\pa}{\pa
\qd^i}\,,\quad{\rm with}\quad i=1,2,3,...
\eeq
and $\lambda$ being an affine parameter. The dot indicates the differentiation with respect to time, $t$. Any function $F(\bq, \bqd)$ is invariant under
the transformation  ${\bf X}$ if \beq \label{06} L_{{\bf
X}}F\eqdef\alp^{i}({\bq})\frac{\pa F}{\pa q^{i}}+
\left(\frac{d}{d\lambda}\alp^{i}({\bq})\right)\frac{\pa F}{\pa
\qd^i}\,=\,0\;, \eeq where $L_{{{\bf X}}}F$ is the Lie derivative
of $F$. Specifically, if $L_{{{\bf X}}}{\cal L}=0$, ${\bf X}$ is a
{\it symmetry} for the dynamics derived by ${\cal L}$. 
What we need is to look for a sufficient condition for which $L_{{{\bf X}}}{\cal L}=0$. The immediate consequence is the {\it Noether Theorem} which states that If $L_{\bf X}{\cal L}=0$, then the function 
\beq \label{09}
\Sigma_{0}=\alp^{i}\frac{\pa {\cal L }}{\pa \qd^i} \,, 
\eeq 
is a constant of motion. As stated in Ref.\cite{j}, the
Noether symmetry approach is a useful tool to select the functions which
assign the models and such functions (and then the models) can be physically relevant. An important question is whether Noether symmetry solutions satisfy the Euler-Lagrange equations of motion or not?. It was demonstrated that on the condition of non-trivial solutions, the Euler- Lagrange and Noether equations are equivalent \cite{A. C. Faliagas}-\cite{TOOBA FEROZE}. By finding any Noether symmetry generator, we will find one associated first integral of the equations of the motion.  Here it gives us a simpler way to integrate the Euler-Lagrange equations of motion emerging from point-like Lagrangian. This conserved charge may be physical like total energy, angular momentum, etc., or a combination of them or just a mathematical expression without any clear physical meaning. 
Furthermore, the existence of Noether symmetries reduce the minisuperspace dimensions and provides a way to make equations of motion integrable \cite{Paliathanasis:2014rja}. Briefly we have two main motivations in studying Noether symmetry approach: The first one is to find conserved charges associated to any symmetry generator and the second one is to make equations of motion integrable and to examine cosmological behaviors of the model. We would suggest the reader to Ref.\cite{j} for further intuitive details.

\subsection{Geometry for Bianchi-I Universe }
The existence of the local anisotropies that we observe today in galaxies, 
cluster and super clusters so at early time can be questioned. 
This may imply that we need something more general than just the isotropy and homogeneous FLRW geometry. In order to go beyond the FLRW universe, we may think of its simplest generalizations. As is well known, the Bianchi Type I model is one of the simplest ones of the anisotropic universe. Unlike FLRW space-time, Bianchi Type I space-time has a different scale factor in each direction, thereby introducing  an anisotropy to the system. The Bianchi Type I line-element in coordinates
$x^{\mu}=(t,x,y,z)$ is
\begin{equation}
ds^2=-dt^2+A^2(t)dx^2+B^2(t)dy^2+C^2(t)dz^2, \label{metric}
\end{equation}
where $A(t),\,B(t)$ and $C(t)$ are the scale factors and they are all functions of the cosmic time $t$. Note that some exact models based on such
metrics have been investigated so far \cite{exact}. Notice that the above metric is just a generalization of a flat FLRW space-time. In the case of the isotropic universe, these variables satisfy $A(t)=B(t)=C(t)$. In other words, the Bianchi Type I model becomes isotropic if the ratio of each directional expansion factor $A(t),\,B(t)$ and $C(t)$ and the expansion factor of the total volume $a(t)$ tends to be a constant value with $a(t)=(ABC)^{1/3}=V^{1/3}$. Here the mean of the three directional Hubble parameters in the Bianchi Type I universe is given by $H=\frac{1}{3}\sum H_i$ where $H_i=\frac{d\ln(A_i)}{dt}$ , $A_i=\{A,B,C\}$. In this paper we are interested in investigating the anisotropic models in which the cosmology is described by the metric (\ref{metric}) with $A\neq B\neq C$. The Bianchi type I model features spatially homogeneous, non-isotropic and non-rotating space-time. In an appropriate coordinate system, they display the diagonal, spatially-Euclidean metric.

\subsection{Point-like Lagrangian and Noether equations}
In this section, we will first write the original Lagrangian in terms of the point-like parameters characterized by the configuration space, i.e. $\mathcal{L}=\mathcal{L}(A,B,C,H_1,H_2,H_3,\dot{H_1},\dot{H_2},\dot{H_3})$. To begin with, we consider the metric (\ref{metric}) and plug into the action (\ref{action}). We then perform integrating by parts to eliminate the terms $\ddot{A}_{i}$, and we
obtain the following point-like Lagrangian, which is suitable for investigation of the symmetry properties of the system:
\begin{eqnarray}\label{la1}
\mathcal{L} &= &\frac{-4}{ABC} \Big(H_{1}^{3}H_{2}+H_{1}^{3}H_{3}+H_{1}^{2}H_{2}^{2}+3\,H_{1}^{2}H_{2}H_{3}+H_{1}^{2}H_{3}^{2}+H_{1}H_{2}^{3} \\&& \nonumber\quad\quad
+3\,H_{1}H_{2}^{2}H_{3}+3\,H_{1}H_{2}H_{3}^{2}+H_{1}H_{3}^{3}
+H_{2}^{3}H_{3}+H_{2}^{2}H_{3}^{2} \\&& \nonumber\quad\quad
+H_{2}H_{3}^{3}+H_{1}^{2}\dot{H_{2}}+H_{1}^{2}\dot{H_{3}}
+H_{1}H_{2}\dot{H}_{1}+H_{1}H_{2}\dot{H_{2}}+2\,H_{1}H_{2}\dot{H_{3}} \\&& \nonumber\quad\quad
+H_{1}H_{3}\dot{H_{1}}+2\,H_{1}H_{3}\dot{H_{2}}+H_{1}H_{3}\dot{H_{3}}
+H_{2}^{2}\dot{H_{1}}+H_{2}^{2}\dot{H_{3}}+2\,H_{2}H_{3}\dot{H_{1}} \\&& \nonumber\quad\quad
+H_{2}H_{3}\dot{H_{2}}+H_{2}H_{3}\dot{H_{3}}+H_{3}^{2}\dot{H_{1}}+H_{3}^{2}\dot{H_{2}}+\dot{H_{2}}\dot{H_{1}}+\dot{H_{3}}\dot{H_{1}}
+\dot{H_{3}}\dot{H_{2}}\Big)\,,
\end{eqnarray}
where we have defined new parameters: 
  \begin{eqnarray}\label{hhh}
 H_1=\frac{\dot{A}}{A}=\frac{d \ln(A(t))}{dt},~~H_2=\frac{\dot{B}}{B}=\frac{d \ln(B(t))}{dt},~~ H_3=\frac{\dot{C}}{C}=\frac{d \ln(C(t))}{dt} .
 \end{eqnarray}
 
Since Nash theory like any other type of the modified theories for gravity designed as a purely geometric model, the role of any type of ordinary or exotic matter is given to the higher-order correction terms in the action and the non trivial (linear) evolution for the corresponding density functions are given in the equations of motion (EoMs) of the model. In our study we just concentrated on the vacuum solutions and studied the role of nonlinear terms in the cosmological evolution. In order to insert the matter field, we can just simply add the matter Lagrangian to the total action and perform metric variation. Note that because still Nash theory is a metric-based theory, the matter energy momentum tensor will be derived as a standard form like other theories.
\par
Note that the unknown functions which must be obtained by this symmetry methods
are $\{H_1,H_2,H_3\}$. We are going to discuss the
Nash cosmology, as a specific case, as follows. As already mentioned, the configuration space is ${\cal Q}=\{H_1,H_2,H_3\}$ while the tangent space
for the related tangent bundle is ${\cal
TQ}=\{H_1,\dot{H}_1,H_2,\dot{H}_2,H_3,\dot{H}_3\}$. The Lagrangian is an application
\beq 
{\cal L}: {\cal TQ}\longrightarrow \Re\,,
\eeq
where $\Re$ is the set of real numbers. We metion here that because Nash gravity is a higher-order derivative theory with respect to the scale factors $A,B,C$, we reduced the difficulty by defining an alternative configuration coordinates set $H_i$ instead of the scale factors. This type of reduction of the configuration coordinastes appeared before in styding Gauss-Bonnet gravity via Noether symmetry as well as many other examples \cite{Capozziello:2014ioa}.
\par
The generator of symmetry in this model reads \beq
\label{vec}{\bf X}=f\frac{\partial}{\partial H_1}+g\frac{\partial}{\partial H_2}+h\frac{\partial}{\partial H_3}+\dot f\frac{\partial}{\partial \dot{H}_1}+\dot g\frac{\partial}{\partial \dot{H}_2}+\dot h\frac{\partial}{\partial \dot{H}_3},.
\eeq 
As discussed above, a symmetry exists if the equation $L_{\bf X}{\cal L}=0$
has solutions. Then there will be a constant of motion on shell,
i.e.\ for the solutions of the Euler equations, as aforementioned.  In other words, a symmetry exists if at
least one of the functions $f,g$ or $h$ in Eq.(\ref{vec})
is different from zero. As a byproduct,  the form of $H_i$ can be determined
in correspondence to such a symmetry. The generalized phase space for our system spans on a a six dimensional manifold with coordinates $(H_a,\dot H_a)$ where $1\leq a\leq 3$,
where $f,g,h$ are functions of $(H_1,H_2,H_3)$ and $\dot{f}=\sum_{a=1}^3\dot{H}_a\frac{\partial f}{\partial {H}_a}$.
The Noether symmetry condition follows
\begin{equation*}
X^{[1]}L =0,
\end{equation*}
which yields the following system of linear PDEs:
\begin{eqnarray}\label{eq1}
&&f   \left( \,{\frac {H_{2}^{2}{\it 
A^\prime}}{A }}+\,{\frac {H_{3}^{2}{\it A^\prime}}{A
  }}-\,H_{2}+\,{\frac {H_{1}\,H_{2}\,{\it A^\prime}}{A
 }}-\,H_{3}+\,{\frac {H_{1}\,H_{3}\,{\it A^\prime}}{A
  }}+2\,{\frac {H_{2}\,H_{3}\,{\it A^\prime}}{A }} \right) \\&& \nonumber 
+g   \left( -2
\,H_{2}+\,{\frac {H_{2}^{2}{\it B^\prime}}{B}}+\,{
\frac {H_{3}^{2}{\it B^\prime}}{B  }}-\,H_{1}+\,{
\frac {H_{1}\,H_{2}\,{\it B^\prime}}{B  }}+\,{\frac {H_
{1}\,H_{3}\,{\it B^\prime}}{B}}-2\,H_{3}+2\,{\frac {H_
{2}\,H_{3}\,{\it B^\prime}}{B }} \right) \\&& \nonumber 
+h \left( \,{\frac {H_{2}^{2}{\it C^\prime}}{C
 }}-2\,H_{3}+\,{\frac {H_{3}^{2}{\it C^\prime}}{C
 }}+\,{\frac {H_{1}\,H_{2}\,{\it C^\prime}}{C  }}-\,H_{1}+\,{\frac {H_{1}\,H_{3}\,{\it C^\prime}}{C }}-2\,H_{2}+2\,{\frac {H_{2}\,H_{3}\,{\it C^\prime}}{C }} \right) \\&& \nonumber 
+ {\frac {\partial f}{\partial H_{1}}}
\left( -\,H_{1}\,H_{2}-
\,H_{1}\,H_{3}-\,H_{2}^{2}-2\,H_{2}\,H_{3}-\,H_{3}^{2} \right) \\&& \nonumber
+  {\frac {\partial g}{\partial H_{1}}} \left( -\,H_{1}^{2}-\,H_{1}\,H_{2}-2\,H_{1}\,H_
{3}-\,H_{2}\,H_{3}-\,H_{3}^{2} \right) \\&& \nonumber
+ {\frac {\partial h} {\partial H_{1}}}  \left( -
\,H_{1}^{2}-2\,H_{1}\,H_{2}-\,H_{1}\,H_{3}-\,H_{2}^{2}-\,H_{2}
\,H_{3} \right)=0\,,
\end{eqnarray}
\begin{eqnarray}
&&\,{\frac {f  {\it A^\prime}}{A }}+\,{\frac {g  {\it B^\prime}}{B
 }}+\,{\frac {h  {\it C^\prime}}{C  }}-\,{\frac {\partial f }{
\partial H_{1}}} -\,{\frac {
\partial g }{\partial H_{2}}}
-\,{
\frac {\partial h}{\partial H_{1}}}  -
\,{\frac {\partial h}{\partial H_{2}}}=0\,,
\end{eqnarray}
\begin{eqnarray}
&&\,{\frac {f  {\it A^\prime}}{A }}+\,{\frac {g  {\it B^\prime}}{B
  }}+\,{\frac {h {\it C^\prime}}{C  }}-\,{\frac {\partial f}{
\partial H_{1}}}  -\,{\frac {
\partial g}{\partial H_{1}}} 
-\,{
\frac {\partial g}{\partial H_{3}}}  -
\,{\frac {\partial h}{\partial H_{3}}}=0\,,
\end{eqnarray}
\begin{eqnarray}
&&\,{\frac {f  {\it A^\prime}}{A  }}+\,{\frac {g  {\it B^\prime}}{B
 }}+\,{\frac {h  {\it C^\prime}}{C }}-\,{\frac {\partial f}{
\partial H_{2}}}  -\,{\frac {
\partial f}{\partial H_{3}}} 
-\,{
\frac {\partial g}{\partial H_{2}}}  -
\,{\frac {\partial h}{\partial H_{3}}}=0\,,
\end{eqnarray}
\begin{eqnarray} \label{eq1t}
&&f   \left( -2\,H_{1}+\,\frac{H_{1}^{2}A^\prime}{A}+\,\frac{H_{3}^{2}A^\prime
}{A}-\,H_{2}+\,\frac{H_{1}\,H_{2}\,A^\prime
}{A}-2\,H_{3}+2\,\frac{H_{1}\,H_{3}\,A^\prime}{A}+\,\frac{H_{2}\,H_{3}\, A^\prime}{A}\right) \\&& \nonumber
 +g \left(\,\frac{H_{1}^{2}B^\prime}{B}+\,\frac{H_{3}^{2}B^\prime}{B}-\,H_{1}+\,\frac{H_{1}\,H_{2}\,B^\prime}{B}+2\,\frac{H_
{1}\,H_{3}\,B^\prime}{B}-\,H_{3}+\,{\frac{H_
{2}\,H_{3}\,{\it B^\prime}}{B  }} \right) \\&& \nonumber 
+h \left( \,\frac{H_{1}^{2}C^\prime}{C
  }-2\,H_{3}+\,\frac{H_{3}^{2}C^\prime}{C
  }+\,\frac{H_{1}\,H_{2}\,C^\prime}{C}-2\,H_{1}+2\,\frac{H_{1}\,H_{3}\, C^\prime}{C}-\,H_{2}+\,\frac{H_{2}\,H_{3}\,C^\prime}{C} \right) \\&& \nonumber 
+ \left(\frac{\partial f}{\partial H_{2}}
 \right) \left( -\,H_{1}\,H_{2}-
\,H_{1}\,H_{3}-\,H_{2}^{2}-2\,H_{2}\,H_{3}-\,H_{3}^{2} \right) \\&& \nonumber 
+ \left(\frac{\partial g}{\partial H_{2}}   \right)  \left( -\,H_{1}^{2}-\,H_{1}\,H_{2}-2\,H_{1}\,H_
{3}-\,H_{2}\,H_{3}-\,H_{3}^{2} \right) \\&& \nonumber
+ \left(\frac{\partial h}{\partial H_{2}}\right)\left( -
\,H_{1}^{2}-2\,H_{1}\,H_{2}-\,H_{1}\,H_{3}-\,H_{2}^{2}-\,H_{2}
\,H_{3} \right)=0\,,
\end{eqnarray}
\begin{eqnarray} \label{eq1g}
&&f   \left( \,\frac{H_{2}^{2}A^\prime}{A}-2\,H_{1}+\,\frac{H_{1}^{2}A^\prime
}{A}-2\,H_{2}+2\,\frac{H_{1}\,H_{2}\,A^\prime
}{A}-\,H_{3}+\,\frac{H_{1}\,H_{3}\,A^\prime
}{A}+\,\frac{H_{2}\,H_{3}\,A^\prime}{A} \right) \\&& \nonumber
 +g \left( -2\,H_{2}+\,\frac{H_{2}^{2}B^\prime}{B}+\,\frac{H_{1}^{2}B^\prime}{B}-2\,H_{1}+ 2\,\frac{H_{1}\,H_{2}\,B^\prime}{B}+
\,\frac{H_{1}\,H_{3}\,B^\prime}{B}-\,H_{3}+
\,\frac{H_{2}\,H_{3}\,B^\prime}{B} \right) \\&& \nonumber
+h \left(\,\frac{H_{2}^{2}C^\prime}{C}+\,\frac{H_{1}^{2}C^\prime}{C
  }+2\,\frac{H_{1}\,H_{2}\,C^\prime}{C}-\,H_{1}+\,\frac{H_{1}\,H_{3}\,C^\prime}{C}-\,H_{2}+\,\frac{H_{2}\,H_{3}\,C^\prime}{C} \right) \\&& \nonumber
+ \left(\frac{\partial f}{\partial H_{3}}\right)  \left( -\,H_{1}\,H_{2}-
\,H_{1}\,H_{3}-\,H_{2}^{2}-2\,H_{2}\,H_{3}-\,H_{3}^{2} \right) \\&& \nonumber 
+ \left(\frac{\partial g}{\partial H_{3}} \right)  \left( -\,H_{1}^{2}-\,H_{1}\,H_{2}-2\,H_{1}\,H_
{3}-\,H_{2}\,H_{3}-\,H_{3}^{2} \right) \\&& \nonumber 
+ \left(\frac{\partial h}{\partial H_{3}}  \right)  \left( -
\,H_{1}^{2}-2\,H_{1}\,H_{2}-\,H_{1}\,H_{3}-\,H_{2}^{2}-\,H_{2}
\,H_{3} \right)=0\,,
\end{eqnarray}
\begin{eqnarray}
{\frac {\partial }{\partial H_{1}}}(g+h)=0 \,,\label{gh}
\end{eqnarray}
and
\begin{eqnarray}
{\frac {\partial }{\partial H_{2}}}(f+h) =0 \,,\label{fh}
\end{eqnarray}
as well as
\begin{eqnarray}
{\frac {\partial }{\partial H_{3}}}(f+g) =0.\label{fg}
\end{eqnarray}
In following section, we are going to figure out particular solutions for the functions $f,g,h$.

\section{General solutions}
\label{IV}
In the following, we are going to solve the system of coupled partial differential equations given in Eqs.(\ref{eq1}-\ref{fg}). Consider, next, Eqs.(\ref{gh}-\ref{fg}). We deduce that:
\begin{eqnarray}
f+g=s(H_{1},H_2),\,\,\,f+h=w(H_{1},H_3),\,\,\,g+h=r(H_2,H_{3}).\label{fgh1}
\end{eqnarray}
where $s,w,r$ are arbitrary functions. From the above expressions, we can obtain
\begin{eqnarray}
&&f = \frac{1}{2}\left(-r(H_2,H_{3}) + s(H_{1},H_2)+ w(H_{1},H_3)\right)\,,\\&& g = \frac{1}{2} \left((r(H_2,H_{3}) + s(H_{1},H_2) - w(H_{1},H_3)\right)\,,\\&&h= \frac{1}{2} \left((r(H_2,H_{3}) - s(H_{1},H_2) + w(H_{1},H_3)\right).\label{fgh}
\end{eqnarray}
Now we are going to solve the above systems (\ref{eq1})-(\ref{fg}) of linear
PDEs. To this end, we take derivative of Eqs.(\ref{gh}-\ref{fg}) and the results can be used to further simplify the last three expressions (\ref{gh})-(\ref{fg}). Here we find 
\begin{eqnarray}
\frac{\d(f+h)}{\d H_1}+\frac{\d}{\d H_2}(g+h)=0 \,,\\
\frac{\d(f+g)}{\d H_1}+\frac{\d}{\d H_3}(g+h)=0 \,,\\
\frac{\d(f+g)}{\d H_2}+\frac{\d}{\d H_3}(f+h)=0 .
\end{eqnarray}
We can simply show that the following equations can be obtained: 
\begin{eqnarray}
\frac{\d}{\d H_1}(g+h)=0 \,,\quad \frac{\d}{\d H_2}(f+h)=0\,, \quad \frac{\d}{\d H_3}(f+g)=0. 
\end{eqnarray}
Consequently we observe that 
\begin{equation}
\frac{\d^2}{\d H_1^2}(f+g)+\frac{\d}{\d H_3}\frac{\d}{\d H_1}(f+g) = \frac{\d^2}{\d H_1^2}(f+g)=0. \label{q1}
\end{equation}
Using the above constraints, after performing an integration on Eq.(\ref{q1}), we come up with the following solutions:
\begin{equation}
f+g=\alpha(H_2) H_1 + \beta(H_2),\label{gen1}
\end{equation}
where $\alpha$ and $\beta$ are functions of $H_{2}$. Similarly we can do the same implementation which allows us to simplify the other equations. Here we find
\begin{equation}
\frac{\d}{\d H_1}\frac{\d}{\d H_2}(f+h) + \frac{\d^2}{\d H_2^2}(g+h) = \frac{\d^2}{\d H_2^2}(g+h)=0\,.
\end{equation}
Again after performing an integration, it yields
\begin{equation}
g+h=\gamma(H_3) H_2 + \delta(H_3),\label{gen2}
\end{equation}
where $\gamma$ and $\delta$ are functions of $H_{3}$. Similarly, we can also obtain the following expression:
\begin{equation}
\frac{\d}{\d H_3}\frac{\d}{\d H_2}(f+g) + \frac{\d^2}{\d H_3^2}(f+h) = \frac{\d^2}{\d H_3^2}(f+h)=0\,,
\end{equation}
for which the solutions read
\begin{equation}
f+h=\theta(H_1) H_3 + \epsilon(H_1),\label{gen3}
\end{equation}
where $\theta$ and $\epsilon$ are functions of $H_{1}$. Using Eqs.(\ref{gen1}), (\ref{gen2}) and (\ref{gen3}) we finally obtain:
\begin{eqnarray}
&&f= \frac{1}{2} \big(\beta(H_2) - \delta(H_3) + \epsilon(H_1) + H_1 \alpha(H_2) - H_2 \gamma(H_3) + H_3 \theta(H_1)\big)\,,\label{ge1}
\\&&
g=\frac{1}{2} \big(\beta(H_2) + \delta(H_3) - \epsilon(H_1) + H_1 \alpha(H_2) +H_2 \gamma(H_3) - H_3 \theta(H_1)\big)\,,\label{ge2}
\\&&
h=\frac{1}{2} \big(-\beta(H_2) + \delta(H_3) + \epsilon(H_1) - H_1 \alpha(H_2) + H_2 \gamma(H_3) + 
   H_3 \theta(H_1)\big).\label{ge3}
\end{eqnarray}
Notice from Eqs.(\ref{ge1})-(\ref{ge3}) that we discover two classes of trivial solutions which can be explicitly written as follows:
\begin{itemize}
\item Class I: We first consider $\alpha=\gamma=\theta=1\,\,\&\,\,\beta=\delta=\epsilon=0$. In this case, the solutions reads
\begin{eqnarray}
&&f= \frac{1}{2} \big(H_1 - H_2 + H_3\big)\,,\label{ge11}
\\&&
g=\frac{1}{2} \big( H_1  +H_2  - H_3 \big)\,,\label{ge21}
\\&&
h=\frac{1}{2} \big( - H_1 + H_2  +  H_3 \big).\label{ge31}
\end{eqnarray}

\item  Class II: The second set of the solutions can be obtained by opting $\alpha=\gamma=\theta=0\,\,\&\,\,\beta=\delta=\epsilon=1$. In this case, we find
\begin{eqnarray}
&&f= \frac{1}{2} \big(H_1 + H_2 -  H_3 \big)\,,\label{ge12}
\\&&
g=\frac{1}{2} \big( - H_1 + H_2 + H_3)\big)\,,\label{ge22}
\\&&
h=\frac{1}{2} \big(H_1 - H_2 +  H_3 \big).\label{ge32}
\end{eqnarray}
\end{itemize}
Notice that in both classes we discover that $2(f+g+h)=H_{1}+H_{2}+H_{3}$. In order to write the general solutions similar to those of Eqs.(\ref{gen1})-(\ref{gen3}), we start rewriting Eqs.(\ref{eq1})-(\ref{eq1g}) as follows:
\begin{eqnarray}\label{w1}
&&f S_1(H_1,H_2,H_3)  
+g  P_1(H_1,H_2,H_3) 
+h Q_1(H_1,H_2,H_3)+\\&&\nonumber
+ {\frac {\partial f}{\partial H_{1}}}
a(H_1,H_2,H_3)
+  {\frac {\partial g}{\partial H_{1}}} b(H_1,H_2,H_3)
+ {\frac {\partial h} {\partial H_{1}}}c(H_1,H_2,H_3)=0\,,
\end{eqnarray}
and
\begin{eqnarray}\label{w2}
&&f S_2(H_1,H_2,H_3)  
+g  P_2(H_1,H_2,H_3) 
+h Q_2(H_1,H_2,H_3)+\\&&\nonumber
+ {\frac {\partial f}{\partial H_{2}}}
a(H_1,H_2,H_3)
+  {\frac {\partial g}{\partial H_{2}}} b(H_1,H_2,H_3)
+ {\frac {\partial h} {\partial H_{2}}}c(H_1,H_2,H_3)=0\,,
\end{eqnarray}
as well as
\begin{eqnarray}\label{w3}
&&f S_3(H_1,H_2,H_3)  
+g  P_3(H_1,H_2,H_3) 
+h Q_3(H_1,H_2,H_3)+\\&&\nonumber
+ {\frac {\partial f}{\partial H_{3}}}
a(H_1,H_2,H_3)
+  {\frac {\partial g}{\partial H_{3}}} b(H_1,H_2,H_3)
+ {\frac {\partial h} {\partial H_{3}}}c(H_1,H_2,H_3)=0\,.\,
\end{eqnarray}
Here we have defined new parameters as follows:
\begin{eqnarray}
&&
S_1(H_1,H_2,H_3)=
(H_2^2+H_3^2+H_1H_2+H_1H_3+2H_2H_3)\frac{A'}{A}-(\,H_{2}+\,H_{3}),\\&&
P_1(H_1,H_2,H_3)=-(H_1+2H_2+2H_3)+(H_2^2+H_3^2+H_1H_2+H_1H_3+2H_2H_3)\frac{B'}{B},\\&&
Q_1(H_1,H_2,H_3)=-(H_1+2H_2+2H_3)+(H_2^2+H_3^2+H_1H_2+H_1H_3+2H_2H_3)\frac{C'}{C},\\&&
S_2(H_1,H_2,H_3)=-(2H_1+H_2+2H_3)+(H_1^2+H_3^2+H_1H_2+H_2H_3+2H_1H_3)\frac{A'}{A},
\end{eqnarray}
\begin{eqnarray}
&&
P_2(H_1,H_2,H_3)=(H_1^2+H_3^2+H_1H_2+H_2H_3+2H_1H_3)\frac{B'}{B}-(\,H_{1}+\,H_{3}),\\&&
Q_2(H_1,H_2,H_3)=-(2H_1+H_2+2H_3)+(H_1^2+H_3^2+H_1H_2+H_2H_3+2H_1H_3)\frac{C'}{C},\\&&
S_3(H_1,H_2,H_3)=-(2H_1+2H_2+H_3)+(H_1^2+H_2^2+2H_1H_2+H_1H_3+H_2H_3)\frac{A'}{A},\\&&
P_3(H_1,H_2,H_3)=-(2H_1+2H_2+H_3)+(H_1^2+H_2^2+2H_1H_2+H_1H_3+H_2H_3)\frac{B'}{B},\\&&
Q_3(H_1,H_2,H_3)=(H_1^2+H_2^2+2H_1H_2+H_1H_3+H_2H_3)\frac{C'}{C}-(\,H_{1}+\,H_{2}),\\&&
a(H_1,H_2,H_3)=  -\,H_{1}\,H_{2}-
\,H_{1}\,H_{3}-\,H_{2}^{2}-2\,H_{2}\,H_{3}-\,H_{3}^{2} ,
\\&&
b(H_1,H_2,H_3)=-\,H_{1}^{2}-\,H_{1}\,H_{2}-2\,H_{1}\,H_
{3}-\,H_{2}\,H_{3}-\,H_{3}^{2},
\\&&
c(H_1,H_2,H_3)= -
\,H_{1}^{2}-2\,H_{1}\,H_{2}-\,H_{1}\,H_{3}-\,H_{2}^{2}-\,H_{2}
\,H_{3} .
\end{eqnarray}
Using the above differential equations, we can mimic the exact solutions. Therefore, they can be generally written as.
\begin{eqnarray}
&&f(H_1,H_2,H_3)=\sum_{i=1}^3 f_i H_i=f_1 H_1+f_2 H_2+f_3 H_3,
\\&&
g(H_1,H_2,H_3)=\sum_{i=1}^3 g_i H_i=g_1 H_1+g_2 H_2+g_3 H_3,
\\&&
h(H_1,H_2,H_3)=\sum_{i=1}^3 h_i H_i=h_1 H_1+h_2 H_2+h_3 H_3,
\end{eqnarray}
where the functions $\{f_i,g_i,h_i\}$ are arbitrary constants (coefficients). In addition, we can also determine the conserved charge, $\Sigma_{0}$, and is given by 
\beq 
\Sigma_{0}= \sum_{i=1}^3 \Big(f_i H_i\frac{\partial\mathcal{L}}{\partial \dot H_1}+  g_i H_i\frac{\partial\mathcal{L}}{\partial \dot H_2}+ h_i H_i\frac{\partial\mathcal{L}}{\partial \dot H_3}\Big)\,,
\eeq
where the point-like Lagrangian, ${\cal L}$, is given in Eq.(\ref{la1}). To be more concrete, we will consider general solutions for two cases of viable scenarios and explicitly determine $f,g,h$ for each scenario. The implementation can be directly done by using Eqs.(\ref{w1}-\ref{w3}). We will investigate the cosmological predictions of these cases  in Sec.({\ref{v}}) 

Scenario I: The first solutions can be simply achieved by adopting $H_{3} = 0, {\dot H}_{3} = 0, H_{2}= {\rm cont.} = m,  {\dot H}_{2} = 0, A = a(t), B= e^{m t}, C= {\rm cont.}$ with $m$ being a constant. After substituting all variables into Eqs.(\ref{w1}-\ref{w3}), we come up with the following solutions:
\begin{eqnarray}
&&f \left( H_{1} \right) ={\frac {\sqrt {H_{1}} \left( 3\,H_{1}+2
\,m \right) {\it C}  }{ \left( {H_{1}}^{2}-H_{1
}\,m-{m}^{2} \right) ^{5/4} }}\,{\rm exp} \left(\frac{3}{10}\,\sqrt{5} \, \, {\rm arctanh}{\left(  \frac{2H_1-m}{m \, \, \sqrt{5}}  \right) }  \right),\\&&g \left( H_{1} \right) =-{\frac {C}{ \sqrt {H_{1}} \left( {H_{1}}^{2}-H_{1}\,m-{m}^{2} \right) ^{5/4} }}\,{\rm exp} \left(\frac{3}{10}\,\sqrt{5} \, \, {\rm arctanh}{\left(  \frac{2H_1-m}{m \, \, \sqrt{5}}  \right) }  \right),\\&&h \left( H_{1} \right) =-{\frac {\sqrt {H_{1}} \left( H_{1}+
\,m \right) {\it C} }{ \left( {H_{1}}^{2}-H_{1}\,m-{m}^{2} \right) ^{5/4} }}\,{\rm exp} \left(\frac{3}{10}\,\sqrt{5} \, \, {\rm arctanh}{\left(  \frac{2H_1-m}{m \, \, \sqrt{5}}  \right) }  \right),
\end{eqnarray}
where $H_{1}$ is already given in Eq.(\ref{c1}). Moreover, by consider the second conditions of parameter space, we can display the following second solutions.

Scenario II: In this second choice, we assume $H_{3} = {\rm cont.}= n, {\dot H}_{3}= 0, H_{2} = {\rm cont.}= m, {\dot H}_{2}= 0, A = a(t), B = e^{m t}, C = e^{n t}$ with $m,\,n$ being constants. After substituting all variables into Eqs.(\ref{w1}-\ref{w3}), we end up with the following solutions:
\begin{eqnarray}
f \left( H_{1} \right) &=& C H_1\left(2n+2m+3H_1 \right)  \nonumber \\
 && \times\,{\rm exp} \left(\frac{1}{2}\,\sum_{i} \, \, {\frac{{\rm ln}(H_1 - R_i) (-6 R_i^2+2(m+n)R_i +m^2+n(6m + n))}{(3 R_i^2-2R_i m -2R_i n-m^2-4 m n -n^2)}  }  \right),
\end{eqnarray}
\begin{eqnarray}
g \left( H_{1} \right) &=& C H_1\left(-H_1^2 + (m-n)H_1 + m(m+n) \right)  \nonumber \\
 && \times\,{\rm exp} \left(\frac{1}{2}\,\sum_{i} \, \, {\frac{{\rm ln}(H_1 - R_i) (-6 R_i^2+2(m+n)R_i +m^2+n(6 m + n))}{(3 R_i^2-2R_i m -2R_i n-m^2-4 m n -n^2)}  }  \right),
\end{eqnarray}
\begin{eqnarray}
g \left( H_{1} \right) &=& C H_1\left(-H_1^2 - (m-n)H_1 + m(m+n) \right)  \nonumber \\
 && \times\,{\rm exp} \left(\frac{1}{2}\,\sum_{i} \, \, {\frac{{\rm ln}(H_1 - R_i) (-6 R_i^2+2(m+n)R_i +m^2+n(6 m + n))}{(3 R_i^2-2R_i m -2R_i n-m^2-4 m n -n^2)}  }  \right),
\end{eqnarray}
where $R_i$'s satisfy the following cubic equation:
\begin{equation}
R_i^3+(m+n)R_i^2+(m^2+4mn+n^2)R_i+mn(m+n)=0,
\end{equation}
It is possible to make integration using EoM and find 
 $H_{1}(t)$ . In the next section we will investigate the cosmological aspects of these two exact viable models.

\section{Cosmological implications}
\label{v}
In this section, we will examine the cosmological implications of the model. Using the above Noether symmetries and the corresponding integral of motions, we can solve Euler-Lagrange equations of motion for Hubble parameters $H_i(t)$. Here the reduction method of superspace using Noether symmetry presented in the Ref.\cite{Paliathanasis:2014rja} is somehow intractable. The reason is that the point-like Lagrangian $\mathcal{L}$ given in Eq.(\ref{la1}) cannot define a simple Kinetic metric. Consequently we devote our study to make integration directly using point-like Lagrangian and will study cosmological parameters based on it. 

In doing that, we start by considering the point-like action given in Eq.(\ref{la1}). To be more concrete, the cosmological implications of the models will be separately discussed in two-case scenarios. Both of these scenarios are based on the general solutions  for the Noether symmetry generators $\{f,g,h\}$ given in the Eqs.(\ref{ge1}-\ref{ge3}). We mention here that the system of Noether equations have many non trivial solutions based on a freedom in functions given in the general forms of the solutions in the Eqs.(\ref{ge1}-\ref{ge3}). We only choice two scenarios among all the possible solutions. In addition, we will show below that each of them predicts cosmological parameters compatible with observational data.   

\subsection{Scenario I}
\begin{figure*}
	\begin{center}
		\includegraphics[width=0.7\linewidth]{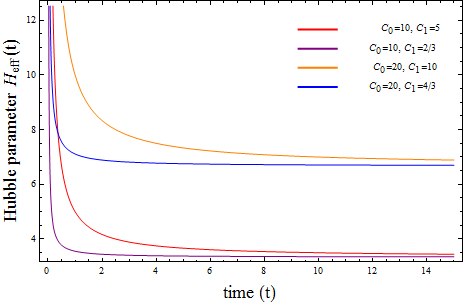}
		\caption{The behavior of the obtained effective Hubble parameter $H_{\rm eff}$ with various values of  $C_{0},\,C_{1}$.} \label{Heff1}
	\end{center}
\end{figure*}

\begin{figure*}
	\begin{center}
		\includegraphics[width=0.47\linewidth]{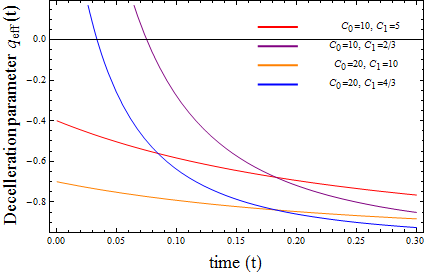}
                     \includegraphics[width=0.48\linewidth]{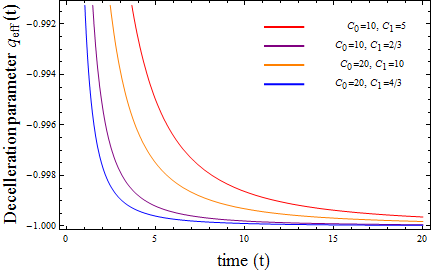}
		\caption{The behavior of the deceleration parameter $q_{\rm eff}(t)$  with various values of  $C_{0},\,C_{1}$ at early time (left panel) and late time (right panel).} \label{qeff}
	\end{center}
\end{figure*}

\begin{figure*}
	\begin{center}
		\includegraphics[width=0.47\linewidth]{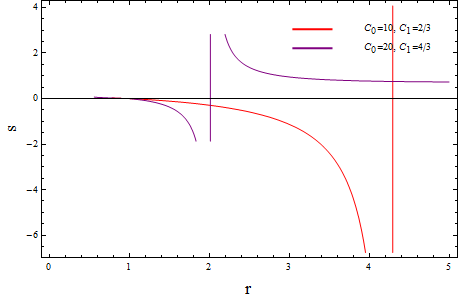}
                     \includegraphics[width=0.48\linewidth]{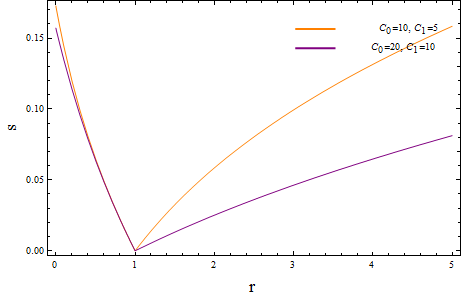}
		\caption{The behavior of the deceleration parameter $q_{\rm eff}(t)$  with various values of  $C_{0},\,C_{1}$ at early time (left panel) and late time (right panel).} \label{rs}
	\end{center}
\end{figure*}

In the first case scenario, we consider the dynamics of $H_{1}$ and derive the equation of motion. Using the action in Eq.(\ref{la1}) together with the assumption $H_{3} = 0, {\dot H}_{3} = 0, H_{2}= {\rm cont.} = m,  {\dot H}_{2} = 0, A = a(t), B= e^{m t}, C= {\rm cont.} = c_{1}$ with $m$ being a constant, we find for $H_{1}$:
\begin{eqnarray}
{\dot H}_{1}(t)+H_{1}(t)^2=0,\quad{\rm with}\quad {\dot{a}(t)=H_{1}a(t)},
\end{eqnarray}
whose solutions take the form
\begin{eqnarray}
H_{1}(t) = \frac{C'_{1}}{t}, \label{c1}
\end{eqnarray}
where $C'_{1}$ is an integrating constant.  In this case, we can define the effective Hubble parameter $H_{\rm eff}$ as
\begin{eqnarray}
H_{\rm eff}(t) \equiv \frac{1}{3}\left(H_{1}+H_{2}+H_{3}\right) = C_{0}+\frac{C_{1}}{t},
\end{eqnarray}
with $C_{0},\,C_{1}$ being constants. The scale factor takes the form $a(t) \propto e^{C_{0}t}t^{C_{1}}$. When $t\ll t_{0}$, in the early Universe and $H_{\rm eff}(t)\sim C_{1}/t$, the Universe was basically filled with perfect fluid; while when $t \gg t_{0}$ the Hubble parameter
$H(t)$ is constant $H_{\rm eff}= C_{0}$ implying that the Universe seems to be de-Sitter. So, this form of $H_{\rm eff}(t)$ provides transition from a matter dominated to the accelerating phase \cite{Nojiri(2006)}. The behavior of the obtained effective Hubble parameter can be displayied in Fig.(\ref{Heff1}) with various values of  $C_{0},\,C_{1}$.
We notice that in all cases the Hubble parameter is positive and becomes constant at the late time. Next the deceleration parameter $q_{\rm eff}(t)$ is defined in terms of the effective Hubble one as follows:
\begin{eqnarray}
q_{\rm eff}(t) \equiv -\left(1 + \frac{{\dot H}_{\rm eff}(t)}{H_{\rm eff}(t)}\right) = -1 + \frac{3 C_{1}}{ \left(C_{1}+C_{0}t\right)^2}.
\end{eqnarray}
The behavior of the deceleration parameter $q_{\rm eff}(t)$ can be clearly illustated in Figs.(\ref{qeff}). Notice that at very early time the deceleration parameters show positivity for rational values of $C_{1}$. However, in all cases, the deceleration parameters are negative and become constants ($\sim -1$) at late time.
Moreover, we can examine other cosmological parameters -- the
statefinder parameters $\{r, s\}$. It is worth noting that statefinder parameters $\{r, s\}=\{1, 0\}$ represents the point where the
flat $\Lambda$CDM model exists in the $r-s$ plane \cite{Huang:2008gs}. So, we can  use this fixed point to test the departure of any particular model from the dark energy ones. We also note that in the $r-s$ plane, a positive value of the parameter $s$ (i.e. $s > 0$) implies a quintessence-like model of dark energy; whilst a negative value of the parameter $s$ (i.e. $s < 0$) indicates a phantom-like model of dark energy. So, different cosmological models, like the models with a cosmological constant $\Lambda$, brane-world
models, chaplygin gas and quintessence models, have been studied using such an analysis
\cite{test}. In this study, it was argued that $\{r, s\}$ can be used to differentiate between different models. Now the parameters can be defined in terms of the effective Hubble parameter as
\begin{eqnarray}
&& r(t) \equiv 1 +3 \frac{{\dot H}_{\rm eff}(t)}{H^{2}_{\rm eff}(t)} + \frac{{\ddot H}_{\rm eff}(t)}{H^{3}_{\rm eff}(t)}=1+\frac{18 C_{1}}{ \left(C_{1}+C_{0}t\right)^3}-\frac{9 C_{1}}{ \left(C_{1}+C_{0}t\right)^2},\label{rt}\\&& s(t) \equiv -\frac{3{\dot H}_{\rm eff}(t)+{\ddot H}_{\rm eff}(t)/H_{\rm eff}(t)}{3\left(2{\dot H}_{\rm eff}(t)+3H^{2}_{\rm eff}(t)\right)} =\frac{1}{C_{1}+C_{0} t}-\frac{C_{0} t}{2 C_{1} C_{0} t+(C_{1}-2)C_{1}+C_{0}^2 t^2}.\label{st}
\end{eqnarray}
We can write $t=t(r)$ by solveing Eq.(\ref{rt}) and substitute $t(r)$ back into Eq.(\ref{st}). Therefore we can obtain $s$ in terms of $r$, i.e., $s=s(r)$. This allow us to make plots the statefinder parameters as displayed in Fig.(\ref{rs})

\subsection{Scenario II}
 In the second case scenario, we consider the dynamics of $H_{1}$ and derive the equation of motion. Using the action in Eq.(\ref{la1}) together with the assumtion $H_{3} = {\rm cont.}= n, {\dot H}_{3}= 0, H_{2} = {\rm cont.}= m, {\dot H}_{2}= 0, A = a(t), B = e^{m t}, C = e^{n t}$ with $m,\,n$ being constants, we find for $H_{1}$:
\begin{eqnarray}
2 \left(m+n\right) H_{1}(t)^2+\left(2 m^2+5mn+2n^2\right) H_{1}(t)+\left(m+n\right)^3 =0,
\end{eqnarray}
whose solutions take the form
\begin{eqnarray}
H^{\pm}_{1}=\frac{-2 m^2-5 mn-2 n^2\pm \sqrt{-4 m^4-12 mn^3-15 n^2 m^2-12 n^3 m-4 n^4}}{4 \left(m+n\right)}= {\rm cont.}\,\,.\label{c2}
\end{eqnarray}
 In this case, the effective Hubble parameter $H_{\rm eff}$ takes a constant value, $M$:
\begin{eqnarray}
H_{\rm eff}(t) \equiv \frac{1}{3}\left(H_{1}+H_{2}+H_{3}\right) = M,
\end{eqnarray}
implying that we have in this case a de-Sitter behavior of the scale factor, i.e. $a(t) \propto e^{Mt}$.  Next the deceleration parameter $q_{\rm eff} = -1$, and the the statefinder parameters $\{r, s\}=\{1, 0\}$ responsible for the flat $\Lambda$CDM model.

\section{Conclusion}

Nash theory of gravity marked one of the alternative theories of gravity. The theory can be viewed as a modification of GR and has been considered to be of interest in attempting to develop theories of quantum gravity. In this paper, we study of Bianchi type-I universe in the context of Nash gravity by using the Noether symmetry approach. We also revisit the Nash theory of gravity. We make a short recap of the Noether symmetry approach and consider the geometry for Bianchi-type I model and consider the geometry for Bianchi-type I model. We obtain the exact general solutions of the theory inherently exhibited by the Noether symmetry. We also examine the cosmological implications of the model by discussing the two cases of viable scenarios. Surprisingly, we find that the predictions are nicely compatible with those of the $\Lambda$CDM model.

However, the solutions we found in the present work have to be further tested with observations. In particular, as presented in \cite{DeLaurentis:2016jfs,Kreisch:2017uet}, scalar-tensor gravity and, in general, Horndeski gravity, can be severely constrained by cosmological and gravitational waves observations. In addition, the astonishingly simple observation has already placed severe constraints on several theories of modified gravity: any modified gravity model predicting $c_T \approx 1$ must now be seriously reconsidered, and several previously viable theories of gravity are now excluded \cite{tests}. Therefore, it is reasonable to use cosmological observations in order to constrain the parameters of Noether symmetries in the Nash theory of gravity.
\section*{Acknowledgements}
We thank  F. M. Mahomed for very useful comments and suggestions.

\end{document}